\documentclass[seceq]{ptptex}

\newcommand{\half}{ \frac{1}{2} }

\newcommand{\bfg}{ \boldsymbol{g} }

\newcommand{\bfA}{ \boldsymbol{A} }

\newcommand{\bfalpha}{\boldsymbol{\alpha}}

\newcommand{\bfeps}{\boldsymbol{\epsilon}}

\newcommand{\bfsig}{\boldsymbol{\sigma}}

\newcommand{\bfomega}{\boldsymbol{\omega}}

\newcommand{\bfGamma}{\boldsymbol{\Gamma}}

\newcommand{\calB}{ {\cal B} }

\newcommand{\calJ}{ {\cal J} }
\newcommand{\calK}{ {\cal K} }

\newcommand{\calP}{ {\cal P} }

\newcommand{\vecd}{ {\vec d} }

\newcommand{\vecp}{ {\vec p} }

\newcommand{\vecv}{ {\vec v} }

\newcommand{\vecx}{ {\vec x} }

\newcommand{\vecA}{ {\vec A} }
\newcommand{\vecB}{ {\vec B} }

\newcommand{\vecE}{ {\vec E} }

\newcommand{\vecJ}{ {\vec J} }

\newcommand{\veccalK}{ {\vec \calK} }
\newcommand{\veccalP}{ {\vec \calP} }

\newcommand{\dotx}{ {\dot x} }

\newcommand{\dotz}{ {\dot z} }

\newcommand{\dotH}{ {\dot H} }

\newcommand{\tilH}{ \tilde{H} }

\newcommand{\tilJ}{ \tilde{J} }

\newcommand{\tilL}{ \tilde{L} }

\newcommand{\tilalpha}{ \tilde{\alpha} }
\newcommand{\tiltheta}{ \tilde{\theta} }

\newcommand{\tilomega}{ \tilde{\omega} }

\newcommand{\Exp}[1]{\left\langle~ #1~ \right\rangle}

\title{%
Hall Effect on Noncommutative Phase Space
}

\author{%
    Akira \textsc{Kokado}$^{1,}$\footnote{
                         E-mail: kokado@kobe-kiu.ac.jp},
    Takashi \textsc{Okamura}$^{2,}$\footnote{
                         E-mail: okamura@ksc.kwansei.ac.jp}
and Takesi \textsc{Saito}$^{3,}$\footnote{
                         E-mail: tsaito@k7.dion.ne.jp}
}
\inst{%
$^1$Kobe International University, Kobe 658-0032, Japan
\\
$^{2,3}$Department of Physics, Kwansei Gakuin University,
    Sanda 669-1337, Japan
}

\abst{%
When phase space coordinates are noncommutative,
especially including arbitrarily noncommutative momenta,
the Hall effect is reinvestigated.
A minimally gauge-invariant coupling of electromagnetic field is
introduced by making use of Faddeev-Jackiw formulation for
unconstrained and constrained systems.
We find that the parameter of noncommutative momenta makes an
important contribution to the Hall conductivity.
}
\begin{document}
\maketitle
\section{Introduction}
Recently the idea of noncommutative structure at small length scales
has been drawn much attention in string theories and field theories
including gravity.
The general expectation was that noncommutative spacetimes could
introduce an effective cut-off in field theories,
in analogy to a lattice. Especially interesting is a model of
open strings propagating in a constant $B$ field background.
Previous studies show that this model is
related to noncommutativity of D-branes,
\cite{rf:1} \
and in the zero slope limit to noncommutative Yang-Mills theory.
\cite{rf:2} \
An intriguing mixing of UV and IR theories has been also found
in the perturbative dynamics of noncommutative field theories.
\cite{rf:3}

In this paper we consider the Hall effect,
when phase space coordinates
\footnote{
For denoting $2N$ dimensional phase space coordinates,
we use upper-case suffix of the Greek letters
$\mu, \nu,\hdots = 1, 2,\hdots, 2N$.
The upper-case small Latin $i, j,\hdots = 1,2, \hdots, N$ are
often used for denoting configuration variables, and
upper-case capital one $I, J,\hdots = 1,2, \hdots, N$ for momenta.
}
are noncommutative.
By the noncommutative phase space we mean the following:
We consider here a system that a particle with charge $e$ moves on
a two-dimensional plane in the uniform external electric field $\vecE$
and the uniform external magnetic field $\vecB$,
where the direction of $\vecE$ is parallel to the $(x^1, x^2)$ plane,
while that of $\vecB$ is transverse to the $(x^1, x^2)$ plane.
Let coordinates of the particle be $\vecx=(x^1, x^2)$ and
momenta be $\vecp=(p^1, p^2)$.
By the noncommutative phase space we mean that
we have a commutation relation in matrix form
\begin{eqnarray}
  \left( \omega^{\mu\nu} \right) & := & \frac{1}{i \hbar}
    \begin{pmatrix}
         [x^1,x^1] & [x^1,x^2] & [x^1,p^1] & [x^1,p^2] \\
         [x^2,x^1] & [x^2,x^2] & [x^2,p^1] & [x^2,p^2] \\
         [p^1,x^1] & [p^1,x^2] & [p^1,p^1] & [p^1,p^2] \\
         [p^2,x^1] & [p^2,x^2] & [p^2,p^1] & [p^2,p^2]
    \end{pmatrix}
   = \begin{pmatrix}
       \theta_x~\bfeps & \bfg^T \\
                - \bfg & \theta_p~\bfeps \end{pmatrix}~,
\label{eq:def-Poisson} \\
  \bfeps &:=& \begin{pmatrix}  0 & 1 \\ -1 & 0 \end{pmatrix}~,
\label{eq:def-bfeps}
\nonumber
\end{eqnarray}
where $\theta_x,~\theta_p$ are real parameters and
$\bfg$ is an arbitrary $2 \times 2$ real matrix.
In this paper we use this commutator together with the Hamiltonian
$H = \vecp\,{}^2/(2m)$.

There have so far been many papers only in the case where
$\theta_x \ne 0,~\theta_p = 0$ and $g^{Ii} = \delta^{Ii}$
(Kronecker delta), together with the Hamiltonian
$H = \vecp\,{}^2/(2m)$.
\cite{rf:4} \
The general type of noncommutativity Eq.(\ref{eq:def-Poisson})
is found in the open string theory with constant $B$ field background.
\cite{rf:5} \
We also know another famous example that the background $B$ field
is so strong as to neglect the kinetic term of a charged particle.
\cite{rf:6} \
In these examples, constraints arise, and then Dirac brackets
can be calculated. As a result, $\theta_x$ and $\theta_p$
are related with each other.

In the present paper we would like to make a step forward such that
$\theta_x$ and $\theta_p$ are generally independent with each other.
Especially, one can see later that the parameter $\theta_p$
of noncommutative momenta makes an important contribution
to the Hall conductivity.
Though one can always make
$[ p^I, p^J ]=i \hbar \theta_p~\epsilon^{IJ}$ to be zero
by means of a congruent transformation,
but in the same time the Hamiltonian is transformed
into another complicated form.
We may take this complicated Hamiltonian
with the transformed commutator $[ p'^I, p'^J ]=0$.
However, this formulation is, of course,
equivalent to the original one. 

Usually the gauge-invariant Hamiltonian is obtained
by the minimal substitution $\vecp~\rightarrow~\vecp - e \vecA(x)$
of electromagnetic field $\vecA(x)$ into the free Hamiltonian.
However, this procedure is valid only when the commutation relations
are canonical, i.e.,
$\theta_x=\theta_p=0$ and $g^{Ii}=\delta^{Ii}$ in
Eq.(\ref{eq:def-Poisson}).
So we should find another procedure of minimal substitution
compatible with Eq.(\ref{eq:def-Poisson}).
For this purpose, in Sec.\ref{sec:gauge}, we follow the line of
Faddeev-Jackiw formulation based on the first-order Lagrangian.
\cite{rf:7}
In Sec.\ref{sec:Hall}, the gauge-invariant coupling thus obtained
is applied to calculate the Hall conductivity
for cases of unconstrained and constrained systems.

The final section is devoted to concluding remarks.
In the Appendix \ref{sec:validity} we discuss the effect of scattering
off a lattice.
In the Appendix \ref{sec:symmetry} we summarize symmetries
and corresponding conserved quantities of our system
with the generalized Poisson bracket.
\section{A gauge-invariant coupling} \label{sec:gauge}
To begin, let us recall the dynamical Hamilton equation of motion
derived from a first-order Lagrangian one-form,
which is usually taken as
\cite{rf:7}
\begin{eqnarray}
  L~dt = \bfalpha (z) - H(z) dt~,
\label{eq:L-one-form}
\end{eqnarray}
where the one-form $\bfalpha(z)=\alpha_\mu(z) dz^\mu$
is an arbitrary function of $z$, and $H(z)$ is the Hamiltonian
which is assumed to be given from the beginning.
Here we have introduced the $2N$-component phase-space coordinate
\begin{eqnarray}
  z^\mu = \begin{cases}
   x^i & \mbox{~~for~} \mu = i = 1, 2, \hdots, N~,
\\
   p^I & \mbox{~~for~} \mu = I + N = N+1, N+2, \hdots, 2N~.
   \end{cases} 
\label{eq:def-variable-z}
\end{eqnarray}
The symplectic two-form $\bfomega(z)$ becomes
\begin{equation}
  \bfomega := d \bfalpha = \half~\omega_{\mu\nu}(z)~
     dz^\mu \wedge dz^\nu~,
\label{eq:def-two-form}
\end{equation}
where 
\begin{equation}
  \omega_{\mu\nu}(z) = \partial_\mu \alpha_\nu
            - \partial_\nu \alpha_\mu~.
\label{eq:two-form-dalpha}
\end{equation}
From the Lagrangian (\ref{eq:L-one-form})
we have the Euler-Lagrange equation 
\begin{eqnarray}
  \omega_{\mu\nu}(z)~\dotz^\nu = \partial_\mu H(z)~.
\label{eq:Hamiltonian-eq}
\end{eqnarray}
When $\omega_{\mu\nu}(z)$ is nonsingular,
the matrix $\omega^{\mu\nu}(z)$ inverse to $\omega_{\mu\nu}(z)$ exists.
Then we have
\begin{eqnarray}
  \dotz^\mu = \omega^{\mu\nu}(z)~\partial_\nu H(z)~,
\label{eq:Hamiltonian-eqI}
\end{eqnarray}
which can be expressed in terms of the generalized Poisson bracket
\begin{eqnarray}
  \dotz^\mu = \{ z^\mu,~H(z) \}~,
\label{eq:Hamiltonian-eq-PB}
\end{eqnarray}
where 
\begin{eqnarray}
  & & \{ z^\mu,~z^\nu \} := \omega^{\mu\nu}(z)
\label{eq:def-FPB}
\end{eqnarray}
with
\begin{eqnarray}
  & &  \{ A,~B \} = \omega^{\mu\nu}(z)~
           \big( \partial_\mu A \big) \big( \partial_\nu B \big)~.
\label{eq:def-PB}
\end{eqnarray}
Hence $\omega_{\mu\nu}(z)$ corresponds to
the generalized Lagrange bracket.

Now, let us introduce minimally the electromagnetic field.
This is achieved by replacement 
\begin{equation}
  H \hspace{0.3cm} \rightarrow \hspace{0.3cm} \tilH = H + e V(x)
\label{eq:def-tilH}
\end{equation}
and
\begin{equation}
  L~dt \hspace{0.3cm} \rightarrow \hspace{0.3cm}
  \tilL~dt = \tilde{\bfalpha}(z) - \tilH(z) dt
\label{eq:def-tilL}
\end{equation}
with $\tilde{\bfalpha}(z):=\bfalpha(z) + e \bfA(z)$,
$\bfA(z) := A_i(x) dx^i$,
where $A_i(x)$ is the vector potential and $V(x)$
is the scalar potential.
It is clear that the above minimal substitution
of the electromagnetic field is gauge-invariant.
According to Eq.(\ref{eq:def-tilL}),
the symplectic two-form (\ref{eq:def-two-form})
is now replaced as
\cite{rf:8}
\begin{equation}
  \bfomega(z) \hspace{0.3cm} \rightarrow \hspace{0.3cm}
  \tilde{\bfomega}(z) = \bfomega(z) + e~d \bfA(z)
  = \bfomega(z) + \half~e~F_{ij}(x)~dx^i \wedge dx^j~,
\label{eq:def-tilde-two-form}
\end{equation}
where $F_{ij}(x) := \partial_i A_j(x) - \partial_j A_i(x)$ is
the field strength. This means that
the generalized Lagrange bracket is replaced as
\begin{equation}
  \omega_{\mu\nu}(z) \hspace{0.3cm} \rightarrow \hspace{0.3cm}
  \tilde{\omega}_{\mu\nu}(z) = 
    \begin{cases} \omega_{ij}(z) + e~F_{ij}(x) &
                       \mbox{~for~} i,j = 1,2,\hdots, N \\
                 \omega_{\mu\nu}(z) & \mbox{~for otherwise}
    \end{cases}~.
\label{eq:def-tilde-two-form-componets}
\end{equation}
Note that $\bfalpha(z)$ is changed into $\tilde{\bfalpha}$ above
by $\bfA(z)$,
but the form of starting Hamiltonian $\tilH(z)$ remains unchanged.
The Hamilton equation of motion is then given by
\begin{eqnarray}
  \dotz^\mu = \{ z^\mu,~\tilH(z) \}~,
\label{eq:Hamiltonian-eq-PB-EM}
\end{eqnarray}
with the generalized Poisson bracket
\begin{eqnarray}
  \{ z^\mu,~z^\nu \} = \tilomega^{\mu\nu}(z)~,
\label{eq:def-PB-EM}
\end{eqnarray}
which is the inverse matrix to $\tilomega_{\mu\nu}(z)$, if it exists.
\section{The Hall effect \label{sec:Hall}}
The above gauge invariant formulation is applied to the Hall effect
in two dimensional configuration space $N=2$.
We start with the generalized Poisson bracket which is given by
\begin{eqnarray}
 \Big(~\{ z^\mu,~z^\nu \}~\Big) = \left(~\omega^{\mu\nu}~\right)
   = \begin{pmatrix} \theta_x~\bfeps & \bfg^T \\
                              - \bfg & \theta_p~\bfeps \end{pmatrix}~,
\hspace{1cm} \mu,~\nu = 1,\hdots, 4~,
\label{eq:def-PoissonI}
\end{eqnarray}
in the same notation as Eq.(\ref{eq:def-Poisson}).
Throughout this paper we assume that its determinant is not zero, i.e.,
$\det( \omega^{\mu\nu} ) = ( \det\bfg - \theta_x \theta_p )^2\ne 0$.
And then, the generalized Lagrange bracket is given by
\begin{eqnarray}
 \left(~\omega_{\mu\nu}~\right) = \frac{1}{D}
   \begin{pmatrix} \theta_p~\bfeps & \bfeps \bfg^T~\bfeps \\
                - \bfeps \bfg \bfeps & \theta_x~\bfeps \end{pmatrix}~,
\hspace{1cm} D := \det\bfg - \theta_x \theta_p \ne 0~.
\label{eq:def-LB}
\end{eqnarray}
We have also the free Hamiltonian given by
\begin{equation}
  H = \frac{1}{2m}~\vecp\,{}^2~, \hspace{1cm}
    \vecp = \begin{pmatrix} p^1 \\ p^2 \end{pmatrix}~.
\label{eq:def-H}
\end{equation}
According to the procedure in Sec.\ref{sec:gauge},
we introduce the magnetic field $F_{ij} = B \epsilon_{ij}$
into $\omega_{ij}$-component
\begin{eqnarray}
 \left(~\tilomega_{\mu\nu}~\right) = \frac{1}{D}~
   \begin{pmatrix} \tiltheta_p~\bfeps & \bfeps \bfg^T~\bfeps \\
                - \bfeps \bfg \bfeps & \theta_x~\bfeps \end{pmatrix}~,
\hspace{1cm} \tiltheta_p := \theta_p + D e B~.
\label{eq:def-LB-EM}
\end{eqnarray}
The inverse matrix to $\tilomega_{\mu\nu}$ is also calculated to be
\begin{eqnarray}
 \left(~\tilomega^{\mu\nu}~\right) = \frac{1}{\lambda}~
   \begin{pmatrix} \theta_x~\bfeps & \bfg^T \\
                         - \bfg & \tiltheta_p~\bfeps \end{pmatrix}~,
\hspace{1cm} \lambda := 1 - \theta_x e B~,
\label{eq:def-FPB-EM}
\end{eqnarray}
where we have also assumed $\lambda \ne 0$.
From Eqs.(\ref{eq:def-LB-EM}) and (\ref{eq:def-FPB-EM}),
we can see that the constant magnetic field
$B$ plays a similar role of the parameter $\theta_p$ of noncommutative
momenta, and vice versa.

According to the replacement (\ref{eq:def-tilH}),
we introduce the electric field $\vecE := ( E_1, E_2 )^T$ by
\begin{equation}
  H \hspace{0.3cm} \rightarrow \hspace{0.3cm}
  \tilH = H + e V(x) = \frac{1}{2m}~\vecp\,{}^2 - e \vecE \cdot \vecx~,
\label{eq:def-H-EM}
\end{equation}
where we have chosen a gauge $V(x)= - \vecE \cdot \vecx$.
In quantum theory, the equation of motion
(\ref{eq:Hamiltonian-eq-PB-EM}) and the generalized Poisson bracket
(\ref{eq:def-PB-EM}) are now replaced by
\begin{eqnarray}
  \dotz^\mu = \frac{1}{i \hbar}~[ z^\mu,~\tilH(z) ]
\label{eq:Heisenberg-eq-EM}
\end{eqnarray}
and
\begin{eqnarray}
  [ z^\mu,~z^\nu ] = i \hbar~\tilomega^{\mu\nu}~,
\label{eq:def-CR-EM}
\end{eqnarray}
respectively.
The equation (\ref{eq:Heisenberg-eq-EM}) reduces to
$\dotz^\mu = \tilomega^{\mu\nu}~\partial_\nu H(z)$, or explicitly
\begin{eqnarray}
 \frac{d}{dt} \begin{pmatrix} \vecx \\ \vecp \end{pmatrix}
  = \frac{1}{\lambda}
   \begin{pmatrix} \theta_x~\bfeps & \bfg^T \\
                         - \bfg & \tiltheta_p~\bfeps \end{pmatrix}
     \begin{pmatrix} - e \vecE \\ \vecp/m \end{pmatrix}~.
\label{eq:Heisenberg-ep-EM-explicit}
\end{eqnarray}
The solution of $\vecp(t)$ is given by
\begin{equation}
  \vecp(t) = \frac{m e}{\tiltheta_p}~\bfeps~\bfg~\vecE
 + \left[~\vecp(0) - \frac{m e}{\tiltheta_p}~\bfeps~\bfg~\vecE~\right]
    e^{ \bfeps (\tilomega_c t) }~,
\label{eq:sol-p}
\end{equation}
with the cyclotron frequency
\begin{equation}
  \tilomega_c := \frac{\tiltheta_p}{m \lambda}
   = \frac{ \theta_p + D e B }{ m ( 1 - \theta_x e B ) }~,
\label{eq:def-tilcyclotron}
\end{equation}
whereas the equation for $\vecx(t)$ is
\begin{equation}
  \frac{d \vecx}{dt} = \frac{1}{\lambda}~\left(
   - \theta_x \bfeps~e \vecE + \frac{1}{m}~\bfg^T~\vecp(t) \right)~.
\label{eq:EOM-vecx-EM}
\end{equation}
Let us consider expectation values of Eqs.(\ref{eq:sol-p})
and (\ref{eq:EOM-vecx-EM}) for a stationary state
of the Hamiltonian $\tilH$.
Since the expectation value $\Exp{\vecp(t)}$ is time-independent,
it follows from Eq.(\ref{eq:sol-p}) that
\begin{equation}
  \Exp{\vecp} = \frac{m e}{\tiltheta_p}~\bfeps~\bfg~\vecE~.
\label{eq:av-p}
\end{equation}
This is substituted into the expectation value of
Eq.(\ref{eq:EOM-vecx-EM}) to obtain
\begin{eqnarray}
  \langle~\dot{\vecx}~\rangle &=&
  = \frac{D e}{ D e B + \theta_p }~\bfeps~\vecE~.
\label{eq:av-dotvecx}
\end{eqnarray}
Here we have used a formula $\bfg^T~\bfeps \bfg = \bfeps (\det\bfg)$.
Hence the expectation value of current
$\langle~\vecJ~\rangle := e \rho~\langle~\dot{\vecx}~\rangle$ with
a charge density $\rho$ is given by
\begin{eqnarray}
  \langle~\vecJ~\rangle = \frac{e \rho}{B}~
     \frac{D e B}{ D e B + \theta_p }~\bfeps~\vecE~.
\label{eq:av-J}
\end{eqnarray}
From this we have the Hall conductivity
\begin{eqnarray}
  \bfsig
   = \frac{e \rho}{B}~\frac{D e B}{ D e B + \theta_p }~\bfeps
   = \frac{e \rho}{B}~\left( 1 - \frac{\theta_p}
          { e B (\det\bfg) + \lambda \theta_p } \right)~\bfeps~,
\label{eq:conductivity}
\end{eqnarray}
where $D = \det\bfg - \theta_x \theta_p$ as in Eq.(\ref{eq:def-LB}).
In the usual noncommutative case with
$\theta_x \ne 0,~\theta_p = 0$ and $\bfg = {\bf 1}$,
the Hall conductivity (\ref{eq:conductivity})
reduces to $\bfsig=( e \rho/B ) \bfeps$, which is equivalent to
the ordinary result in the commutative case with
$\theta_x = \theta_p = 0$ and $\bfg = {\bf 1}$ and has no effect
from any parameters of noncommutativity.

Thus we see that the parameter $\theta_p$ makes
an important contribution to the Hall conductivity.
This is apparent from the fact that the parameter $\theta_p$
plays a role of an effective magnetic field.
By using the effective magnetic field $B_p:=\theta_p/(eD)$,
the Hall conductivity is also written as
\begin{eqnarray}
  \bfsig
   = \frac{e \rho}{B}~\frac{1}{1 + B_p/B }~\bfeps~. 
\label{eq:conductivityII}
\end{eqnarray}

It is worthwhile to comment the role of $\theta_x$ on the cyclotron
frequency $\tilomega_c$ given by Eq.(\ref{eq:def-tilcyclotron}).
By the existence of $\theta_x$, the dependence of the cyclotron
frequency on the magnetic field $B$ is qualitatively different from
that in commutative case $\theta_x=0$, such that at the limit of
strong field $|B| \gg 1/|e \theta_x|$, $\tilomega_c$ tends to
a finite value $\tilomega_c \rightarrow - D/(m \theta_x)$ and
$\tilomega_c$ is singular at $B= 1/(e \theta_x)$.

\vspace*{0.3cm}
{\bf A singular case ($\lambda = 0$)}

\noindent
The determinant of $\tilomega_{\mu\nu}$ in Eq.(\ref{eq:def-LB-EM})
is given by
\begin{equation}
  \det( \tilomega_{\mu\nu} ) = \frac{ 1 - \theta_x e B }
   { \det\bfg - \theta_x \theta_p } = \frac{\lambda}{D}~.
\label{eq:det-LB-EM}
\end{equation}
When $\lambda = 1 - \theta_x e B = 0$,
there is no inverse matrix of $\tilomega_{\mu\nu}$,
so that constraints arise.
In this case the parameter $\theta_x$ is given by $\theta_x= 1/(eB)$,
and then $\tilomega_{\mu\nu}$ reduces to
\begin{eqnarray}
 \big(~\tilomega_{\mu\nu}~\big) = \frac{1}{D}
   \begin{pmatrix}  e B (\det\bfg)~\bfeps & \bfeps \bfg^T~\bfeps \\
                - \bfeps \bfg \bfeps & 1/(eB)~\bfeps \end{pmatrix}~.
\label{eq:def-LB-EM-degenerate}
\end{eqnarray}
The variation of Lagrangian $L=\tilalpha_\mu \dotz^\mu - \tilH(z)$
yields the equation of motion
\begin{eqnarray}
  \tilomega_{\mu\nu}~\dotz^\nu =  \partial_\mu \tilH(z)~,
\label{eq:Hamiltonian-eq-degenerate}
\end{eqnarray}
which reduces to
\begin{eqnarray}
  \frac{1}{D}~
   \begin{pmatrix}  e B (\det\bfg)~\bfeps & \bfeps \bfg^T~\bfeps \\
                - \bfeps \bfg \bfeps & 1/(eB)~\bfeps \end{pmatrix}
   \begin{pmatrix} \dot{\vecx} \\ \dot{\vecp} \end{pmatrix}
 = \begin{pmatrix} - e \vecE \\ \vecp/m \end{pmatrix}~.
\label{eq:def-LB-EM-degenerate-explicit}
\end{eqnarray}
Multiplying the matrix
\begin{eqnarray}
  \begin{pmatrix} {\bf 1} & {\bf 0} \\
                - {\bf 1} & e B \bfeps \bfg^T \end{pmatrix}
\label{eq:def-Multiplying}
\end{eqnarray}
to Eq.(\ref{eq:def-LB-EM-degenerate-explicit}),
we obtain the equation of motions
\begin{equation}
  e B~( \det\bfg )~\bfeps~\dot{\vecx}
    + \bfeps \bfg^T~\bfeps~\dot{\vecp} = - D~e \vecE~,
\label{eq:EOM-degenerate}
\end{equation}
and constraints
\begin{equation}
  \vecE = - \frac{B}{m}~\bfeps \bfg^T~\vecp~,
\label{eq:constraint-degenerate}
\end{equation}
that is, $\vecp$ becomes a constant vector.
This reflects the fact that
the matrix Eq.(\ref{eq:def-LB-EM-degenerate})
has a rank two and two zero eigenvalues.
Hence we can drop the second term on the L.H.S.
in Eq.(\ref{eq:EOM-degenerate}) to obtain
\begin{equation}
  e \rho~\dot{\vecx}
 = \frac{e \rho}{B}~\frac{D}{\det\bfg}~\bfeps \vecE~.
\label{eq:J-degenerate}
\end{equation}
By taking an expectation value for Eq.(\ref{eq:J-degenerate})
we obtain the Hall conductivity
\begin{equation}
  \bfsig = \frac{e \rho}{B}~\frac{D}{\det\bfg}~\bfeps
  = \frac{e \rho}{B}~\left( 1 - \frac{\theta_p}{eB \det\bfg} \right)~
    \bfeps~.
\label{eq:conductivity-degenerate}
\end{equation}
This result coincides with the value obtained by formal substitution
$\lambda=0$ into Eq.(\ref{eq:conductivity}).
From this result we see again that the parameter $\theta_p$
makes an important contribution to the Hall conductivity.
\section{Concluding remarks}
We have considered the Hall effect when phase space coordinates
are noncommutative, especially including the parameter
$\theta_p$ of noncommutative momenta.
First a gauge-invariant coupling of electromagnetic field
has been considered when the generalized Lagrange bracket
is given by $\omega^{\mu\nu}$ in Eq.(\ref{eq:def-Poisson}).
This bracket corresponds to noncommutativity of the phase space.
In order to find such a coupling,
it is convenient to consider the one form of Faddeev-Jackiw
$\bfalpha(z)$ in Eq.(\ref{eq:L-one-form}),
which is a generalization of conjugate momentum.
The vector potential one form $\bfA(x)$ is then substituted into
$\bfalpha(z)$ as $\bfalpha(z) \rightarrow \bfalpha(z) + e \bfA(x)$.
This is our gauge invariant minimal substitution of the gauge field.
According to this substitution, the generalized Lagrange bracket
is changed into $\tilomega^{\mu\nu}(z)$ in Eq.(\ref{eq:def-PB-EM}).
This formulation is applied to the Hall effect.
A calculation based on $\tilomega^{\mu\nu}(z)$ in
Eq.(\ref{eq:def-LB-EM}) leads to the result Eq.(\ref{eq:conductivity}).
We also considered a singular case where there exists
no generalized Lagrange bracket and constraints arise.
The result is given in Eq.(\ref{eq:conductivity-degenerate}).
We have found that the noncommutativity parameters,
especially $\theta_p$, make an important contribution
to the Hall conductivity,
because the parameter $\theta_p$ plays a role of
a constant homogeneous magnetic field.

In this paper we have neglected effects from many particle correlations
and also from scattering off a lattice.
The former approximation may be justified
when the density of charged particle is low.
For the latter it may be justified when the magnetic field $B$ is
so chosen that the inequality
\begin{equation}
  \left\vert~\frac{e}{m}~\left( B + \frac{\theta_p}{eD} \right)~
       \right\vert \gg \frac{1}{\tau}~,
\label{eq:validity}
\end{equation}
holds (see Appendix \ref{sec:validity}), where
$\tau$ is the relaxation time due to the scattering.
One can see that there is a very interesting region near
$B = 1/(\theta_x e)$ satisfying this condition.

In the Appendix \ref{sec:symmetry} we have summarized symmetries
and corresponding conserved quantities of our system
with the generalized Poisson bracket (\ref{eq:def-PoissonI}) and
the Hamiltonian (\ref{eq:def-H}).
These symmetries are associated with time-translation, rotation,
space translation and Galilei boost.
%
\section*{Acknowledgements}
We are grateful to Reijiro Kubo and Gaku Konisi
for their critical comments.
\appendix
\section{Derivation of the inequality (\ref{eq:validity})}
\label{sec:validity}
We take into account of an dissipation by interaction with a lattice,
and derive the inequality (\ref{eq:validity}) as a condition for
neglecting this effect.

In order to incorporate the dissipation phenomenologically,
we add environmental degrees of freedom to our system.
After integrating out the equation of motion of environment,
we obtain an effective one for our system with dissipation as
\cite{rf:9}
\begin{eqnarray}
  & & \omega_{\mu\nu} \dotz^\nu(t) = \partial_\mu H(z)
        + \delta_\mu{}^i~\delta V_{ij}~~x^j(t)
        + \delta_\mu{}^i~\frac{m}{\tau}~\Gamma_{ij}~\dotx^j(t)
        + \cdots~,
\label{eq:effective-EOMII}
\end{eqnarray}
where $\delta V_{ij}$ is a constant,
$\tau$ is a relaxation time and $\Gamma_{ij}$ are non-dimensional
$O(1)$ quantities.
The second term of R.H.S. in Eq.(\ref{eq:effective-EOMII}) shows
the potential renormalization effect and hereafter we discard
the second term by using renormalized Hamiltonian $H_R$.
The third term stands for \lq\lq Ohmic\rq\rq~dissipation effect,
and it is incorporated by the modification of Lagrange bracket
\begin{eqnarray}
  \omega_{\mu\nu} \hspace{0.3cm} \longrightarrow \hspace{0.3cm}
  \omega^D_{\mu\nu} = \omega_{\mu\nu}
       - \delta_\mu{}^i~\frac{m}{\tau}~\Gamma_{ij}~\delta^j{}_\nu~.
\label{eq:def-omegaD}
\end{eqnarray}
The effective equation of motion is then written as
\begin{eqnarray}
  & & \omega^D_{\mu\nu}~\dotz^\nu(t) = \partial_\mu H_R(z) + \cdots~.
\label{eq:EOM-with-disspation}
\end{eqnarray}
If $\det( \omega^D_{\mu\nu} ) \ne 0$, we have the modified
Poisson bracket $\omega_D^{\mu\nu}$ including the dissipation,
which is defined as the inverse matrix to $\omega^D_{\mu\nu}$.
Then Eq.(\ref{eq:EOM-with-disspation}) can be rewritten as
\begin{eqnarray}
  & & \dotz^\mu = \omega_D^{\mu\nu}~\partial_\nu H_R(z) + \cdots~.
\label{eq:EOM-with-disspationII}
\end{eqnarray}

In general, the modified Lagrange bracket $\omega^D_{\mu\nu}$
and the corresponding Poisson bracket $\omega_D^{\mu\nu}$ cannot be
anti-symmetric owing to the dissipation $\Gamma_{\mu\nu}$.
Indeed, only those symmetric parts cause energy dissipation and
phase volume contraction, according to equations
\begin{eqnarray}
  & & \dotH_R = \dotz^\mu~\partial_\mu H_R + \cdots
  = \omega_D^{(\mu\nu)}~\left( \partial_\mu H_R \right)~
    \left( \partial_\nu H_R \right) + \cdots~,
\label{eq:energy-disspation} \\
  & & \partial_\mu \dotz^\mu
     = \omega_D^{(\mu\nu)}~\partial_\mu \partial_\nu H_R + \cdots~.
\label{eq:volume-disspation}
\end{eqnarray}

When the system couples to the electromagnetic field in such a way that
the generalized Lagrange bracket and the Hamiltonian are given by
Eqs.(\ref{eq:def-LB-EM}) and (\ref{eq:def-H-EM}), respectively,
the modified Lagrange bracket and the equation of motion
due to the dissipation become as follows:
\begin{eqnarray}
  & & \tilomega^D_{\mu\nu} = \tilomega_{\mu\nu}
       - \delta_\mu{}^i~\frac{m}{\tau}~\Gamma_{ij}~\delta^j{}_\nu~,
\label{eq:def-omegaD-EM} \\
  & & \tilomega^D_{\mu\nu}~\dotz^\nu(t) = \partial_\mu \tilH_R(z)
        + \cdots~.
\label{eq:EOM-with-disspation-EM}
\end{eqnarray}
The matrix notation of Eq.(\ref{eq:def-omegaD-EM}) is written by
\begin{eqnarray}
 \left(~\tilomega^D_{\mu\nu}~\right) = \frac{1}{D}~
   \begin{pmatrix} \tiltheta_p~\bfeps - m~D \bfGamma/\tau
                                        & \bfeps \bfg^T~\bfeps \\
              - \bfeps \bfg \bfeps & \theta_x~\bfeps \end{pmatrix}~,
\hspace{1cm} \bfGamma = ( \Gamma_{ij} )~.
\label{eq:def-LB-EM-dissipation}
\end{eqnarray}
Therefore, we conclude that we can neglect the dissipation effect
due to scattering off a lattice, if the magnetic field $B$ is so chosen
that the inequality $|\tiltheta_p| \gg m|D|/\tau$, or equivalently,
\begin{equation}
  \left\vert~\frac{e}{m}~\left( B + \frac{\theta_p}{eD} \right)~
       \right\vert \gg \frac{1}{\tau}~,
\label{eq:validity-Append}
\end{equation}
holds.

When $\lambda \ne 0$, the cyclotron frequency is given by
$\tilomega_c = \tiltheta_p/(m \lambda)$. Hence this inequality is also
written as
\begin{eqnarray}
  & & \left\vert~\tilomega_c \tau~\right\vert \gg
  \left\vert~\frac{D}{\lambda}~\right\vert
  = \left\vert~\frac{\det\bfg - \theta_x \theta_p}{1 - \theta_x e B}~
      \right\vert~.
\label{eq:validity-nodegenerate}
\end{eqnarray}
%
\section{Symmetries of the system with the generalized Poisson Bracket}
\label{sec:symmetry}
We summarize symmetries of the system
with the generalized Poisson bracket (\ref{eq:def-PoissonI})
and the Hamiltonian (\ref{eq:def-H}).
These symmetries are associated with 1) time translation,
2) rotation , 3) space translation and Galilei boost.

We can make use of the Noether's theorem on the phase space:
The variation of an action
\begin{eqnarray}
 S[z] = \int^{t_f}_{t_i} dt~L \big(~z,~\dot{z},~t~\big)
      = \int^{t_f}_{t_i} dt~\big[~\alpha_\mu (z)~\dotz^\mu
            - H(z)~\big]~,
\label{eq:action}
\end{eqnarray}
under the infinitesimal variations
\begin{eqnarray}
  t \,\, \rightarrow \,\, t'= t + \delta t(t)~,
\hspace{1cm}
  z^\mu (t) \,\, \rightarrow \,\,
     z'^\mu (t') = z^\mu(t) + \delta z^\mu(t)~,
\label{eq:tran}
\end{eqnarray}
becomes
\begin{eqnarray}
  & & \delta S[z] = - \big[~J(t)~\big]^{t_f}_{t_i}~,
\hspace{1cm}
  J(t) := - \delta t~L - \delta_L z^\mu(t)~
           \frac{\partial L}{\partial \dotz^\mu}~,
\label{eq:def-J}
\end{eqnarray}
where $\delta_L z^\mu$ is the Lie variation, $~\delta_L z^\mu (t)
:= z'^\mu(t) - z^\mu(t) = \delta z^\mu(t) - \delta t~\dotz^\mu(t)$ and
we use the equation of motion.

Applying the above results to our Lagrangian
\begin{eqnarray}
  & & L \big( z,~\dot{z} \big)
   = \half~z^\mu \omega_{\mu\nu}~\dotz^\nu - H(z)~,
\label{eq:simple-Lag}
\end{eqnarray}
where $\omega_{\mu\nu}$ and $H(z)$ are given by
Eqs.(\ref{eq:def-LB}) and (\ref{eq:def-H}), respectively,
we can show conserved quantities associated with
the above four symmetries.

\vspace*{0.5cm}
%
\noindent
{\bf 1)~Time translation:} \\
With an arbitrarily infinitesimal constant $\eta$,
the time translation is given by
\begin{eqnarray}
  & & \delta t = \eta~,
\hspace{1.5cm}
      \delta z = 0 = \delta_L z^\mu + \eta~\dotz^\mu~,
\label{eq:time-tr}
\end{eqnarray}
under which the action is invariant.
Therefore the corresponding conserved quantity is
\begin{eqnarray}
  & & J = \eta~\left(~
      \dotz^\mu \frac{\partial L}{\partial \dotz^\mu} - L~\right) 
   = \eta~H(z)~.
\label{eq:time-tr-J}
\end{eqnarray}
Namely, our Hamiltonian is the conserved quantity just as
in usual systems.

\vspace*{0.3cm}
%
\noindent
{\bf 2)~Rotation:} \\
With an arbitrarily infinitesimal constant $\eta$,
we consider variations
\begin{eqnarray}
  & & \delta t = 0~, \hspace{1.5cm} 
  \delta z^\mu = \delta_L z^\mu := - \eta~\Lambda^\mu{}_\nu~z^\nu~,
\label{eq:rotation-tr} \\
  & & ( \Lambda^\mu{}_\nu ) := - \frac{1}{\det\bfg}~
    \begin{pmatrix} \bfg^T~\bfg \bfeps & {\bf 0} \\
          {\bf 0} & \left( \det\bfg \right)~\bfeps \end{pmatrix}~,
\label{eq:def-Lambda}
\end{eqnarray}
under which the action is invariant.
Where $\Lambda^\mu{}_\nu$ holds
$\omega_{\mu\lambda} \Lambda^\lambda{}_\nu = 
\omega_{\nu\lambda} \Lambda^\lambda{}_\mu$.

The corresponding conserved quantity is given by
\begin{eqnarray}
  \calJ &:=& \eta^{-1}~J
  = \half~\omega_{\mu\rho}~\Lambda^\rho{}_\nu~z^\mu z^\nu
\nonumber \\
  &=& \frac{\theta_p}{~2~D \det\bfg \,}~
       \left\vert~\bfg~\bfeps~\vecx~\right\vert^2
  + \frac{1}{D}~\vecx^T~\bfeps~\bfg^T~\vecp
  + \frac{\theta_x}{2 D}~\left\vert~\vecp~\right\vert^2~.
\label{eq:def-calJ}
\end{eqnarray}
In the usual Poisson bracket,
the above variations reduce to $\delta \vecx
= \eta~\bfeps~\vecx$ and $\delta \vecp = \eta~\bfeps~\vecp$,
representing an infinitesimal rotation,
and $\calJ$ becomes the usual angular momentum
$\calJ = \vecx^T~\bfeps~\vecp$.

\vspace*{0.3cm}
%
\noindent
{\bf 3)~space translation and Galilei boost:} \\
With an arbitrarily infinitesimal constant vector
$( \zeta^\mu ) := ( d^i,~m v^I )$, we consider variations
\begin{eqnarray}
  & & \delta t = 0~, \hspace{1.5cm} 
     \delta z^\mu = \delta_L z^\mu := - \Gamma^\mu{}_\nu \zeta^\nu~,
\label{eq:parallel-Galilei-tr} \\
  & & ( \Gamma^\mu{}_\nu )
  := \begin{pmatrix} {\bf 1} & - \theta_p^{-1}~\bfg^T \bfeps~
      \Big[~{\bf 1} - \exp\big\{ (\theta_p t/m)~\bfeps \big\}~\Big] \\
    {\bf 0} & - \exp\big\{~(\theta_p t/m)~\bfeps~\big\} \end{pmatrix}~.
\label{eq:def-Gamma}
\end{eqnarray}
Where $\Gamma^\mu{}_\nu$ holds
$\Lambda^\mu{}_\lambda~\Gamma^\lambda{}_\nu =
\Gamma^\mu{}_\lambda~\Lambda^\lambda{}_\nu$ and
$\Gamma^\rho{}_\mu~\omega_{\rho\sigma}~\Gamma^\sigma{}_\nu
= \omega_{\mu\nu}$.

Under the variation Eq.(\ref{eq:parallel-Galilei-tr}), we have
\begin{eqnarray}
  \delta S[z] &=& \int^{t_f}_{t_i} dt~
    \half~\delta z^\mu~\omega_{\mu\nu}~\dotz^\nu = \Big[~
    \half~\delta z^\mu~\omega_{\mu\nu}~z^\nu~\Big]^{t_f}_{t_i}~.
\label{eq:parallel-Gallilei-S-tr}
\end{eqnarray}
Though the action is not invariant owing to the surface term,
but one can define a new current
\begin{eqnarray}
  \tilJ &:=& J(t) + \half~\delta z^\mu~\omega_{\mu\nu}~z^\nu
   = \delta z^\mu~\omega_{\mu\nu}~z^\nu
   = \zeta^\mu~\left( z^\nu~\omega_{\nu\lambda}~
      \Gamma^\lambda{}_\mu \right)~,
\label{eq:def-tilJ}
\end{eqnarray}
which is conserved. 

Since $\zeta^\mu$ is arbitrary, we have the conserved quantities
$\calB_\mu = ( \calP_i,~\calK_I ) :=
z^\nu~\omega_{\nu\lambda}~\Gamma^\lambda{}_\mu$, or in vector notation
\begin{eqnarray}
  & & \veccalP := - \frac{ \theta_p~\bfeps~\vecx
        + \bfeps~\bfg^T~\bfeps~\vecp }{D}~,
\label{eq:def-veccalP} \\
  & & \veccalK := \frac{ \bfeps^{-1} \bfg \bfeps }{D}~\vecx
  - \frac{ \sin\left( \theta_p~t/m \right) }{ \theta_p/m }~
       \frac{\vecp}{m}
  + \left[~ \frac{m \theta_x}{D}
     + \frac{ 1 - \cos\left( \theta_p~t/m \right) }{ \theta_p/m }~
    \right]~\bfeps~\frac{\vecp}{m}~.
\label{eq:def-veccalK}
\end{eqnarray}
In the usual Poisson bracket,
with $\theta_x=\theta_p=0$ and $\bfg={\bf 1}$,
the above variations reduce to
$\delta \vecx = - \vecd + \vecv~t$ and $\delta \vecp = m \vecv$,
representing infinitesimal translation and Galilei boost,
and $\veccalP$ and $\veccalK$ become $\vecp$ and $\vecx - \vecp~t/m$,
respectively.

It is easy to show that those conserved quantities
are generators of the corresponding variations
\begin{eqnarray}
  & & \left\{ z^\mu,~\eta~\calJ \right\}
  = \eta~\Lambda^\mu{}_\nu~z^\nu = - \delta_L z^\mu~,
\label{eq:generatorI} \\
  & & \left\{ z^\mu,~\zeta^\nu~\calB_\nu \right\}
  = \Gamma^\mu{}_\nu~\zeta^\nu = - \delta_L z^\mu~.
\label{eq:generatorII}
\end{eqnarray}

And then, it is worth to note that the conserved quantities holds
the algebraic relations as
\begin{eqnarray}
  & & \left\{ \calJ,~\calB_\mu \right\}
  = \calB_\nu~\Lambda^\nu{}_\mu~,
\label{eq:algebraI} \\
  & & \left\{ \calB_\mu,~\calB_\nu \right\}
  = - \Gamma^\rho{}_\mu~\omega_{\rho\sigma}~\Gamma^\sigma{}_\nu
  = - \omega_{\mu\nu}~,
\label{eq:algebraII}
\end{eqnarray}
that is homomorphic to the two parameters extension of
two-dimensional isochronous Galilei Lie algebra.
\cite{rf:10}


\begin{thebibliography}{99}
\bibitem{rf:1}
   C. Chu and P. Ho, \NPB{550,1999,151}, hep-th/9812219.\\
   M. M. Sheikh-Jabbari, \PLB{455,1999,129}, hep-th/9901080.\\
   F. Ardfaei and M. M. Sheikh-Jabbari, \JHEP{02,2000,016},
             hep-th/9810072; \NPB{576,2000,578}, hep-th/9906161.
\bibitem{rf:2} 
   N. Seiberg and E. Witten, \JHEP{09,1999,032}, hep-th/9908142.
\bibitem{rf:3}
   S. Minwalla, M. Van Raamsdonk and N. Seiberg, \JHEP{02,2000,020},
             hep-th/9912072.\\
   A. Matusis, L. Susskind and N. Toumbas, \JHEP{12,2000,002},
             hep-th/0002075.
\bibitem{rf:4} 
   S. Bellucci, A. Nersessian and C. Sochichiu, \PLB{522,2001,345},
      hep-th/0106138. \\
   S. Bellicci and A. Nersessian, \PLB{542,2002,295}, hep-th/0205024. \\
   C. Duval and P. A. Horvathy, \PLB{479,2000,284}, hep-th/0002233;~
       \JP{A34,2001,10097}, hep-th/0106089. \\
   P. A. Horvathy, \ANN{299,2002,128}. \\
   V. P. Nair and A. P. Polychronakos, \PLB{505,2001,267},
      hep-th/0011172. \\
   B. Morariu and A. P. Polychronakos, \NPB{610,2001,531},
      hep-th/0102157;~ibid. \andvol {634,2002,326}. \\
   O.F. Dayi and A. Jellal, \JMP{43,2002,4592}, hep-th/0111267. \\
   F. J. Ezawa, G. Tsitsishvilli and K. Hasebe, \lq\lq
     Noncommutative Geometry, Extended Algebra and Grassmannian
       Solitons in Multicomponent Quantum Hall Systems \rq\rq,
         hep-th/0209198.
\bibitem{rf:5}
   A. Kokado, G. Konisi and T. Saito, \PTP{104,2000,1289}, 
             hep-th/0009190.
\bibitem{rf:6}
   T. Saito, Gravitation and Cosmology {\bf 6}, No.22(22), (2000), 130.
\\
   G. Dune and R. Jackiw, Nucl. Phys. B(proc. Suppl.)
      \andvol{33C,1993,114}. \\
   D. Bigatti and L. Susskind, hep-th/9908142.
\bibitem{rf:7}
   L. Faddeev and R. Jackiw, \PRL{60,1988,1692}.
\bibitem{rf:8}
   For references, see also Horvathy, et.al. in \cite{rf:4}.
\bibitem{rf:9}
   For discussions on systems with the ordinary Poisson bracket,
   see the references, for instance,
   A.O. Caldeira and A.J. Leggett, \PRA{31,1985,1059}. \\
   B.L. Hu, J.P. Paz and Y. Zhang, \PRD{45,1992,2843}. \\
   T. Okamura, \PTP{91,1994,219}. \\
   J.J. Halliwell and T. Yu, \PRD{53,1996,2012}.
\bibitem{rf:10}
   J.-M. L\'evy-Leblond,
      \lq\lq Galilei group and Galilean invariance\rq\rq, in
   \textit{Group theory and Applications, Vol. II}, ed. E. Loebl
    (Academic Press, New York,1972).
\end{thebibliography}
\end{document}